\documentclass{article}

\usepackage{arxiv}

\usepackage[utf8]{inputenc} 
\usepackage[T1]{fontenc}    
\usepackage{hyperref}       
\usepackage{url}            
\usepackage{booktabs}       
\usepackage{amsfonts}       
\usepackage{nicefrac}       
\usepackage{microtype}      
\usepackage{lipsum}		
\usepackage{graphicx}
\usepackage{natbib}
\usepackage{doi}

\usepackage{subcaption}
\usepackage{amsmath}
\usepackage{multirow}
\usepackage{placeins}
\usepackage{graphicx}

\newcommand{\orcid}[1]{%
  \href{https://orcid.org/#1}{%
    \includegraphics[height=1.6ex]{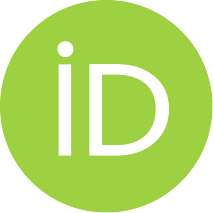}%
  }%
}

\title{The 2026 Algorithmic Information Theory Data Compression Challenge}

\author{%
\begin{minipage}{0.95\textwidth}
\centering
André Ribeiro\orcid{0009-0008-0194-7171} $^{1}$, Rúben Garrido\orcid{0009-0002-1508-7885}$^{1}$, Violeta Ramos\orcid{0009-0005-9803-4982}$^{1}$,
António Alberto\orcid{0009-0007-9201-2840}$^{1}$, Diogo Fernandes\orcid{0009-0009-1395-179X}$^{1}$, João Varela\orcid{0009-0000-9578-1054}$^{1}$,
Eduardo Lopes\orcid{0009-0009-5054-4886}$^{1}$, Rodrigo Abreu\orcid{0009-0004-8629-4130}$^{1}$, Hugo Ribeiro\orcid{0009-0009-5912-5602}$^{1}$,
Tomás Brás\orcid{0009-0006-7503-7926}$^{1}$, David Pelicano\orcid{0009-0008-3642-1364}$^{1}$, Afonso Ferreira\orcid{0009-0003-6454-9334}$^{1}$,
Sebastião Teixeira\orcid{0009-0006-4663-1387}$^{1}$, Maria Linhares\orcid{0009-0004-8630-8432}$^{1}$, Martim Santos\orcid{0009-0002-1444-9258}$^{1}$,
Rui Machado\orcid{0009-0005-7757-6619}$^{1}$, Duarte Santos\orcid{0009-0009-4696-5514}$^{1}$, Gabriel Silva\orcid{0009-0003-8792-6331}$^{1}$,
Guilherme Rosa\orcid{0009-0005-9220-9665}$^{1}$, João Roldão\orcid{0009-0006-5204-5712}$^{1}$, Henrique Teixeira\orcid{0009-0005-2160-2718}$^{1}$,
Cláudia Seabra\orcid{0009-0008-3156-0963}$^{1}$, Ricardo Fonseca\orcid{0009-0004-0390-5371}$^{1}$, Richard Miranda\orcid{0009-0006-2283-6396}$^{1}$,
Hugo Castro\orcid{0009-0005-2884-8534}$^{1}$, Ângela Ribeiro\orcid{0009-0006-3273-3606}$^{1}$, Fouad Bellili\orcid{0009-0005-8637-0671}$^{1}$,
Luís Diogo\orcid{0009-0007-7332-8970}$^{1}$, André Cardoso\orcid{0009-0006-8736-2320}$^{1}$, Armando J. Pinho\orcid{0000-0002-9164-0016}$^{1,2,3,*}$,
and Diogo Pratas\orcid{0000-0003-1176-552X}$^{1,2,3,4,*}$
\end{minipage}
\\[0.5em]
{\small\begin{minipage}{\linewidth}\begin{center}
\begin{tabular}{ccc}
$^1$DETI - Department of Electronics, Telecommunications and Informatics,\\ University of Aveiro, 3810-193 Aveiro, Portugal \\
$^2$IEETA - Institute of Electronics and Informatics Engineering of Aveiro, and\\
$^3$LASI - Intelligent Systems Associate Laboratory, and \\
$^4$DoV - Department of Virology, University of Helsinki, 00014 Helsinki, Finland\\
$^*$Corresponding authors: ap@ua.pt and pratas@ua.pt\\
\end{tabular}
\end{center}\end{minipage}}
}

\date{}

\renewcommand{\shorttitle}{The 2026 Algorithmic Information Theory Data Compression Challenge}

\hypersetup{
pdftitle={The 2026 Algorithmic Information Theory Data Compression Challenge},
pdfsubject={The 2026 AITDCC},
pdfauthor={André Ribeiro, Rúben Garrido, Violeta Ramos, António Alberto, Diogo Fernandes, João Varela, Eduardo Lopes, Rodrigo Abreu, Hugo Ribeiro, Tomás Brás, David Pelicano, Afonso Ferreira, Sebastião Teixeira, Maria Linhares, Martim Santos, Rui Machado, Duarte Santos, Gabriel Silva, Guilherme Rosa, João Roldão, Henrique Teixeira, Cláudia Seabra, Ricardo Fonseca, Richard Miranda, Hugo Castro, Ângela Ribeiro, Fouad Bellili, Luís Diogo, André Cardoso, Armando J. Pinho, Diogo Pratas},
pdfkeywords={Data Compression, Algorithmic Information Theory},
}

\begin{document}
\maketitle

\begin{abstract}
Lossless data compression remains central to computer science, with direct impact on storage, communication bandwidth, computational cost, and energy consumption. It is also closely related to Algorithmic Information Theory, where compressibility provides an operational measure of structure and non-randomness. This paper presents the 2026 Algorithmic Information Theory Data Compression Challenge, a benchmark for evaluating general-purpose lossless compressors under realistic constraints. Submissions were encouraged to use arithmetic or range coding, limited to at most 8 GB of memory, and required to include a decompressor no larger than 1 MB. The benchmark comprised sixteen heterogeneous files, split into public training and hidden testing datasets. In total, 117 valid submitted compressors were evaluated alongside established reference compressors using compression ratio, compression and decompression time, Weissman score, and Pareto-frontier analysis. The results show that performance depends strongly on the optimization criterion: fast compressors achieved the best speed-oriented scores, whereas modelling-intensive compressors produced smaller outputs at higher computational cost. A Normalized Compression Distance analysis further revealed clusters of related submissions and distinguished incremental variants from more independent implementations. Selected submissions were described for their methodological novelty or competitive performance and further tested on four large external datasets, where several achieved competitive or superior results relative to established compressors. Overall, the challenge confirms the importance of probabilistic modelling, hidden testing, and external datasets for assessing compression performance and generalization. Benchmark resources, leaderboard data, binaries, and selected source code are publicly available at \url{https://aitdcc.github.io}.
\end{abstract}


\section{Introduction}
\label{sec:intro}

Data compression is a fundamental component of modern digital infrastructures because it reduces the volume of data that must be stored, transmitted, and processed, thereby lowering storage costs, bandwidth demands, and energy consumption in large-scale computing systems. Its importance is therefore not only technological and economic, but also environmental, and improvements in compression efficiency are aligned with broader sustainability goals, including the United Nations Sustainable Development Goals related to infrastructure, responsible resource use, and climate action \cite{un2015sdgs}. At a deeper level, compression is linked to finding structure in data: a sequence can be compressed only when it is not random and contains regularities that an algorithm can capture \cite{salomon2007data,sayood2017introduction}. This connection between compression, regularity, and computational structure is also consistent with Wolfram's broader computational perspective on simple rules, complexity, and data compression \cite{wolfram2002newkind}. This naturally connects practical compression with Algorithmic Information Theory \cite{chaitin1977algorithmic}, where compressibility is understood as an operational indicator of structure and non-randomness \cite{kolmogorov1968three,chaitin1975randomness,solomonoff1964formal,li2019kolmogorov}.

Motivated by this connection between compression, structure, and algorithmic information, the 2026 Algorithmic Information Theory Data Compression Challenge provided the experimental framework for the present work. The challenge focused on the design of general-purpose lossless compressors under realistic implementation constraints: participants submitted both a compressor and a decompressor, the maximum memory (Random Access Memory) footprint was limited to 8~GB, and the decompressor size was limited to 1~MB \cite{aitproject2026,aitdcc2026}. The benchmark was organized into eight public training files and eight hidden (during the competition) testing files, thereby discouraging direct optimization for the final evaluation data and emphasizing generalization to unseen inputs \cite{aitdcc2026}. The datasets were deliberately heterogeneous, comprising protein sequences, source code, Wikipedia text, pseudo-random sequences, high-energy physics numerical data, astronomical raw images, and executable binaries. This diversity made the challenge a robust experimental setting for evaluating the adaptability of general-purpose compressors across markedly different statistical and structural regimes.

Within this framework, arithmetic coding or an equivalent range coder was strongly encouraged. This design choice is relevant because modern lossless compressors can be viewed as the combination of two tightly coupled components: a modelling stage, which estimates symbol probabilities from the data, and a coding stage, which converts those probabilities into compact binary representations \cite{salomon2007data,witten1987arithmetic}. Arithmetic coding is particularly suitable for this setting because it encodes symbols according to the probabilities assigned by the model and can approach the ideal code length more closely than coding schemes restricted to integer-length codewords \cite{witten1987arithmetic}. Thus, its use in the challenge was important not only from the viewpoint of coding efficiency, but also because it promoted a clear separation between probabilistic modelling and entropy coding. This separation is scientifically useful. When the coding stage is fixed or standardized, performance differences can be more directly attributed to modelling choices, such as context selection, probability estimation, adaptation mechanisms, or the combination of multiple predictors. Moreover, the probabilistic model can be analyzed independently of the final binary representation and reused as an instrument for detecting structure in data or even to generate it \cite{shannon1951prediction,sagan2025family,pratas2011dna}. This view is consistent with Wolfram’s description of language generation as an iterative process in which a model repeatedly selects plausible continuations from the text generated so far \cite{wolfram2023chatgpt}. This perspective is particularly relevant when compression is used not only to reduce file size, but also to quantify regularity, similarity, complexity, or change.

More generally, compression models and compression-based distances provide a domain-independent way to quantify structure, similarity, complexity, and change in data. Because they operate directly on symbolic or binary representations, they have been used to compare and cluster heterogeneous objects such as texts, languages, music, images, astronomical data, software, and biological sequences without requiring handcrafted features \cite{bennett1998information,li2004similarity,cilibrasi2005clustering,nikvand2019perceptually,vazquez2012using}. In software analysis, digital forensics, and cybersecurity, related approaches can reveal byte-level regularities associated with file-fragment structure, shared code, packing effects, or anomalous organization \cite{axelsson2010ncd,alshahwan2015malware}. In biological sequence analysis, the same principles support alignment-free genomics and proteomics, including the characterization of sequence complexity \cite{silva2022complexity}, identification of low-complexity regions in genomic and proteomic sequences \cite{silva2023alcor}, analysis of temporal variation in sequence collections \cite{silva2024altair}, and detection of genomic rearrangements \cite{hosseini2020smashpp}. Thus, compression is not only a storage mechanism, but also a quantitative approach for studying organization, similarity, and structure across heterogeneous data types.

From a broader computational perspective, these modelling principles must also be considered alongside the practical trade-offs that define general-purpose lossless compression. State-of-the-art compressors span a wide speed--memory--compression design space. Classical dictionary- and transform-based tools such as \texttt{gzip}, \texttt{bzip2}, and \texttt{xz} remain important baselines for portability, robustness, and compression performance \cite{deutsch1996deflate,deutsch1996gzip,burrows1994block,seward2019bzip2,pavlov2026lzma,collin2024xz}. More recent systems such as \texttt{LZ4}, \texttt{zstd}, and Brotli prioritize high throughput, efficient decompression, and scalable deployment in real-time, storage, and web-oriented settings \cite{collet_lz4,collet2021zstandard,alakuijala2019brotli,rfc7932}. At the high-compression end, the PAQ-family compressors remain strong references for predictive modelling and context mixing, although they usually incur substantially higher computational costs and are increasingly challenged by learning-based compression approaches \cite{knoll2011paq8,mahoney2024paq,goyal2018deepzip,farhat2025dicomp,ma2025msdzip,sun2025survey}.

The online challenge evaluation reported standard performance indicators, including compression ratio, compression time, and decompression time, and also included the Weissman score, computed against \texttt{gzip} for the selected dataset \cite{aitdcc2026}. In total, 117 valid compressors were submitted, reflecting an extensive process of iterative refinement in modelling. Several submissions achieved competitive performance relative to strong reference compressors, and some surpassed reference methods under specific benchmark conditions. Other submissions proved particularly effective for certain classes of data, despite being designed as general-purpose methods. These observations are especially relevant because no single compressor dominates all practical scenarios: fast codecs such as \texttt{brotli} and \texttt{zstd} are often preferred when throughput is critical, whereas \texttt{xz}, or PAQ-like methods may be more appropriate when compression ratio is prioritized over encoding speed.

Overall, the AIT data compression challenge provides a valuable experimental setting for studying practical lossless compression under realistic resource constraints, such as those imposed by contemporary laptops. Moreover, it offers an opportunity to examine how general-purpose compressors can be improved through modelling choices, how performance varies across heterogeneous data types, and how benchmark design can promote genuine generalization rather than dataset-specific tuning. Since the submitted implementations and benchmark resources were made publicly available, the challenge also constitutes a useful contribution to the wider data compression community.

The remainder of this article is organized as follows. Section~\ref{sec:methodology} describes the benchmark dataset, submission interface, implementation constraints, execution protocol, evaluation metrics, Pareto-frontier analysis, binary similarity analysis, reference compressors, selected submitted methodologies, external dataset and computational environment. Section~\ref{sec:results} presents the experimental results, including the Pareto analysis, Weissman-score ranking, compressed-size ranking, binary similarity heatmap, and external evaluation on selected heterogeneous datasets. Finally, Section~\ref{sec:conclusion} summarizes the main findings.


\section{Methodology}
\label{sec:methodology}

This study evaluates the 2026 Algorithmic Information Theory Data Compression Challenge as a benchmark for general-purpose lossless compression under practical implementation constraints. The methodology covers the benchmark data, submission requirements, execution protocol, performance metrics, Pareto-frontier analysis, leaderboard generation, external dataset, description of the selected methodologies, and computational environment.

\subsection{Benchmark Dataset}

The benchmark consisted of sixteen fixed files, labeled A--P, divided into a public \emph{training} partition and a hidden \emph{testing} partition:
\begin{equation}
S_{\mathrm{train}} = \{A,B,C,D,E,F,G,H\},
\qquad
S_{\mathrm{test}} = \{I,J,K,L,M,N,O,P\}.
\end{equation}
The training files were available to the participants during the competition and could be used for development, tuning, and comparison through the online leaderboard. The testing files were withheld until the final evaluation, reducing the possibility of direct optimization for the complete benchmark and allowing the final results to better reflect generalization to unseen data.

The contents of the benchmark files are summarized in Table~\ref{tab:benchmark_dataset}. The training and testing partitions were constructed to cover related but distinct data regimes. Thus, for several data classes, the testing file was not a duplicate of the corresponding training file, but a different instance of the same broad source type.

\begin{table}[!h]
\centering
\caption{Benchmark datasets: A--H formed training; I--P formed hidden testing.}
\label{tab:benchmark_dataset}
{\small
\setlength{\tabcolsep}{5pt}
\renewcommand{\arraystretch}{1.05}
\begin{tabular}{c c l}
\hline
File & Partition & Description \\
\hline
A & Training & Enterococcus phage protein sequence data \\
B & Training & C source code \\
C & Training & English Wikipedia text \\
D & Training & Pseudo-random sequence \\
E & Training & Floating-point numerical data from CERN ATLAS \\
F & Training & Raw astronomical image data \\
G & Training & Raw astronomical image data \\
H & Training & Executable binary file (zstd) \\
\hline
I & Testing & Human protein sequence data \\
J & Testing & Source code-derived data from the \texttt{zlib} project \\
K & Testing & English Wikipedia text segment distinct from file C \\
L & Testing & Pseudo-random sequence generated with a different seed \\
M & Testing & Floating-point numerical data from CERN ATLAS \\
N & Testing & Raw astronomical image data \\
O & Testing & Raw astronomical image data \\
P & Testing & Binary package data (deb) \\
\hline
\end{tabular}
}
\end{table}

This composition was deliberately heterogeneous. Protein sequences, natural language, and source code contain symbolic regularities and repeated motifs; scientific numerical data contain structured floating-point patterns; raw astronomical images contain spatial and instrumental redundancy; pseudo-random data provide a near-incompressible control case (if parameters are unknown); and executable or packaged binary data contain compiled-code structure, metadata, alignment effects, and machine-level regularities. Consequently, strong performance on the benchmark required compressors to adapt to multiple statistical regimes rather than exploit a single dominant source of redundancy.

\subsection{Submission Interface and Constraints}

Each submission consisted of a lossless compression system with a compressor and a decompressor. To ensure uniform evaluation, submissions followed a common command-line interface:
\begin{equation}
\texttt{CompressorName inputFileName outputFileName},
\end{equation}
\begin{equation}
\texttt{DecompressorName inputFileName outputFileName}.
\end{equation}
Wrapper scripts were used when needed to adapt individual implementations to this interface.

The challenge imposed two main resource constraints:
\begin{equation}
M_{\mathrm{peak}} \leq 8~\mathrm{GB},
\qquad
B_{\mathrm{decomp}} \leq 1024~\mathrm{kB},
\end{equation}
where \(M_{\mathrm{peak}}\) denotes peak memory usage and \(B_{\mathrm{decomp}}\) denotes the decompressor binary size. Participants were also strongly encouraged to use arithmetic coding or an equivalent range-coding method as the entropy-coding stage. These constraints favoured compact, executable, and practically deployable compressors rather than unconstrained offline models.

For compressor \(c\) and file \(i\), a result was considered valid only if compression completed successfully, decompression completed successfully, the reconstructed file was produced, and the reconstruction was byte-identical to the original. This validity condition can be expressed as
\begin{equation}
L_{c,i}
=
\mathbb{I}
\left[
\mathrm{SHA256}(X_i)
=
\mathrm{SHA256}(\widehat{X}_{c,i})
\right],
\end{equation}
where \(X_i\) is the original file, \(\widehat{X}_{c,i}\) is the decompressed reconstruction, and \(L_{c,i}=1\) denotes a valid lossless result. Failed compression, failed decompression, missing outputs, or reconstruction mismatches were marked as non-lossless.

\subsection{Benchmark Execution Protocol}

The benchmark was implemented as a Bash-based evaluation pipeline. For each compressor and dataset file, the script compressed the original file, decompressed the compressed representation, and verified the reconstructed output. Compressed size was measured with \texttt{stat}, losslessness was verified with \texttt{sha256sum}, and compression and decompression wall-clock times were measured with \texttt{/usr/bin/time}. Optional CPU pinning through \texttt{taskset} was supported, although the benchmark was designed to remain portable on a standard Linux system.

To reduce the effect of occasional runtime fluctuations, each compression and decompression command was executed three times in the reported configuration, and the median wall-clock time was retained:
\begin{equation}
T^{\mathrm{comp}}_{c,i}
=
\operatorname{median}
\left(
t^{\mathrm{comp}}_{c,i,1},
t^{\mathrm{comp}}_{c,i,2},
t^{\mathrm{comp}}_{c,i,3}
\right),
\end{equation}
\begin{equation}
T^{\mathrm{decomp}}_{c,i}
=
\operatorname{median}
\left(
t^{\mathrm{decomp}}_{c,i,1},
t^{\mathrm{decomp}}_{c,i,2},
t^{\mathrm{decomp}}_{c,i,3}
\right).
\end{equation}
For each compressor \(c\) and file \(i\), the benchmark therefore recorded the original size \(O_i\), compressed size \(C_{c,i}\), median compression time \(T^{\mathrm{comp}}_{c,i}\), median decompression time \(T^{\mathrm{decomp}}_{c,i}\), and losslessness indicator \(L_{c,i}\).

The total runtime for compressor \(c\) on file \(i\) was defined as
\begin{equation}
T^{\mathrm{total}}_{c,i}
=
T^{\mathrm{comp}}_{c,i}
+
T^{\mathrm{decomp}}_{c,i}.
\end{equation}

For a set of files \(S\), such as the training, testing, or overall partition, sizes and runtimes were aggregated by summation:
\begin{equation}
O_S = \sum_{i \in S} O_i,
\qquad
C_{c,S} = \sum_{i \in S} C_{c,i},
\end{equation}
\begin{equation}
T^{\mathrm{comp}}_{c,S}
=
\sum_{i \in S} T^{\mathrm{comp}}_{c,i},
\qquad
T^{\mathrm{decomp}}_{c,S}
=
\sum_{i \in S} T^{\mathrm{decomp}}_{c,i},
\end{equation}
\begin{equation}
T^{\mathrm{total}}_{c,S}
=
T^{\mathrm{comp}}_{c,S}
+
T^{\mathrm{decomp}}_{c,S}.
\end{equation}
A compressor was considered lossless on an aggregate partition only if it reconstructed every file in that partition as
\begin{equation}
L_{c,S}
=
\prod_{i \in S} L_{c,i}.
\end{equation}
Thus, a single failed file was sufficient to mark the corresponding aggregate result as non-lossless.

\subsection{Performance Metrics}

The primary measure of compression effectiveness was compression ratio. For compressor \(c\) on file set \(S\), the compression ratio was computed as
\begin{equation}
R_{c,S}
=
\frac{O_S}{C_{c,S}},
\end{equation}
and the compression ratio in bits per original byte was computed as
\begin{equation}
\mathrm{bpb}_{c,S}
=
8 \frac{C_{c,S}}{O_S}.
\end{equation}
Lower values of \(C_{c,S}\) and \(\mathrm{bpb}_{c,S}\), and higher values of \(R_{c,S}\), indicate stronger compression. As described in the previous subsection, runtime was reported separately as compression time, decompression time, and total time, allowing compression effectiveness to be assessed together with computational cost.

The online leaderboard also reported a Weissman score using \texttt{gzip} as the reference compressor \cite{weissman_score}. Let \(R_{g,S}\) and \(T^{\mathrm{total}}_{g,S}\) denote the compression ratio and total time of \texttt{gzip}, and let \(R_{c,S}\) and \(T^{\mathrm{total}}_{c,S}\) denote the corresponding values for compressor \(c\). The score was computed as
\begin{equation}
W_{c,S}
=
\frac{R_{c,S}}{R_{g,S}}
\cdot
\frac{\log(T^{\mathrm{total}}_{g,S}+1)}
{\log(T^{\mathrm{total}}_{c,S}+1)}.
\end{equation}
This metric rewards compression-space improvements relative to \texttt{gzip} while penalizing high runtime. It should therefore be interpreted as a normalized space--speed trade-off rather than as a pure compressed-size ranking.

\subsection{Pareto-Frontier Analysis}

Pareto frontiers were computed for the training, testing, and overall partitions using two objectives: total execution time and compressed size in bits per original byte. Since both quantities are minimized, a valid compressor \(c\) was considered dominated on file set \(S\) if there existed another valid compressor \(d\) such that
\begin{equation}
T^{\mathrm{total}}_{d,S}
\leq
T^{\mathrm{total}}_{c,S}
\quad
\text{and}
\quad
\mathrm{bpb}_{d,S}
\leq
\mathrm{bpb}_{c,S},
\end{equation}
with at least one inequality strict as
\begin{equation}
T^{\mathrm{total}}_{d,S}
<
T^{\mathrm{total}}_{c,S}
\quad
\text{or}
\quad
\mathrm{bpb}_{d,S}
<
\mathrm{bpb}_{c,S}.
\end{equation}
Equivalently, \(d\) dominated \(c\) if it was no slower and no larger, and was strictly better in at least one objective.

The Pareto frontier for partition \(S\) was defined as the set of valid, non-dominated compressors as
\begin{equation}
\mathcal{P}_S
=
\left\{
c :
L_{c,S}=1
\;\land\;
\not\exists d
\text{ such that }
d \prec_S c
\right\},
\end{equation}
where \(d \prec_S c\) denotes domination on file set \(S\). Non-lossless compressors were excluded from this analysis. In the Pareto plots, non-dominated compressors were highlighted with larger numbered markers and connected by a black line, while all valid compressors were shown as background points. The horizontal axis reports \(T^{\mathrm{total}}_{c,S}\), and the vertical axis reports \(\mathrm{bpb}_{c,S}\); therefore, points closer to the lower-left region represent better speed--compression trade-offs.

\subsection{Binary Similarity Analysis}
\label{subsec:ncd_method}

The Normalized Compression Distance (NCD) was used to quantify binary-level similarity between compressor executables in an algorithm-agnostic manner. The purpose of this analysis was to determine whether different compressor submissions shared implementation structure, common code lineage, or incremental changes across successive versions.

Pairwise similarity between compressor binaries was estimated using NCD, a practical compression-based approximation to the theoretical Normalized Information Distance (NID), which is derived from Kolmogorov complexity \cite{bennett1998information,li2004similarity,cilibrasi2005clustering}. For two binary files \(x\) and \(y\), and a lossless compressor \(Z\), the distance was computed as
\begin{equation}
\mathrm{NCD}_{Z}(x,y)
=
\frac{
Z(xy)-\min\{Z(x),Z(y)\}
}{
\max\{Z(x),Z(y)\}
},
\label{eq:ncd}
\end{equation}
where \(Z(x)\) and \(Z(y)\) are the compressed sizes of the individual binaries, and \(Z(xy)\) is the compressed size of their concatenation. In this work, \(Z\) was computed using Zstandard in mode 15 (\texttt{zstd-15} where the -15 suffix stands for mode/level) as the underlying lossless compressor \cite{collet2021zstandard}. The resulting pairwise distances were assembled into a square matrix,
\begin{equation}
\mathbf{D}
=
\left[
\mathrm{NCD}_{Z}(x_i,x_j)
\right]_{i,j=1}^{n},
\end{equation}
where \(n\) is the number of evaluated compressor binaries and \(x_i\) denotes the \(i\)-th binary. The diagonal entries were set to zero, since each binary is identical to itself:
\begin{equation}
\mathrm{NCD}_{Z}(x_i,x_i)=0.
\end{equation}

Lower NCD values indicate greater shared compressible structure between two binaries and, consequently, higher binary-level similarity under the selected compressor. Conversely, values close to one indicate little shared structure. Therefore, in the heatmap representation, compact low-distance blocks suggest closely related implementations, common code bases, or incremental version updates, whereas uniformly high distances suggest independent implementations or substantially different binary structure. 

\subsection{Reference Compressors and Submissions}

The benchmark included established reference compressors, for example \texttt{gzip}, \texttt{bzip2}, \texttt{zstd}, \texttt{lzma}, and \texttt{brotli}, with several of them evaluated at multiple compression levels. Additional high-compression references, including PAQ-family compressors, were included to contextualize the submitted methods against strong modelling-based approaches.

For reference compressors, suffixes were generally appended to the compressor names to indicate the selected compression level or mode. For example, `lzma-9` denotes `lzma` running at compression level 9, equivalent to invoking `lzma -9`.

Submissions (from groups) were organized by group (G) and version (V), for example \texttt{G1-V1}, \texttt{G1-V2}, and subsequent variants. Multiple versions from the same group reflected iterative development during the challenge, including changes in modelling strategy, context selection, parameter tuning, coding method, and implementation details. This organization made it possible to analyze both final compressor performance and the evolution of compressor design across versions.

\subsection{Leaderboard Generation}

The benchmark script produced two comma-separated output files. The file \texttt{results\_groups.csv} contained the complete benchmark output, including dataset label, compressor name, original size, compressed size, compression time, decompression time, losslessness flag, baseline flag, and plotting color. The file \texttt{results\_for\_html.csv} contained the subset of fields required by the public leaderboard.

The online leaderboard was implemented as a static HTML page with an embedded CSV data block. Client-side JavaScript parsed the CSV data, computed derived quantities such as \(R_{c,S}\), \(\mathrm{bpb}_{c,S}\), \(T^{\mathrm{total}}_{c,S}\), rank, and \(W_{c,S}\), and rendered sortable tables and interactive plots. Results could be inspected for individual files, the training partition, the testing partition, and the overall aggregate as described in the benchmark execution protocol. Training and testing rows were computed by summing the corresponding individual-file results, whereas the overall view used the explicit aggregate rows produced by the benchmark script.

The leaderboard also included filtering options to select or exclude compressors by name, with multiple entries specified as comma-separated strings. This allowed users to focus the tables and plots on selected submissions, compressor families, or reference methods.

This design made the benchmark transparent and reproducible, since the raw numerical data, ranking logic, aggregation rules, plotting procedure, and Weissman-score computation were visible in the public leaderboard source.

\subsection{Selected Compression Methodologies}
\label{subsec:selected_methods}

At their core, most submitted compression tools--from the 117 valid compressors--combined several established lossless-compression techniques, including the Burrows-Wheeler Transform (BWT) \cite{burrows1994block}, Move-To-Front (MTF) coding \cite{ryabko1980data}, Run-Length Encoding (RLE) \cite{robinson1967results}, context-based statistical models, and arithmetic coding. The main novelty therefore did not usually come from entirely new primitives, but from how these components were organized and adapted. In particular, several approaches introduced preprocessing stages, file-type classification, transform-selection mechanisms, model mixing, parallelization approaches, and alternative orderings of the compression pipeline to better match the statistical structure of each input.

Among the submitted approaches, we selected six representative compression methodologies: G1-V22, G2-V3, G5-V25, G7-V8, G7-V10, G9-V1. These compressors were chosen based on their methodological distinctiveness, relevance to the Pareto-frontier results, Weissman score, and position in the binary similarity analysis presented in the Results section. Together, they represent competitive submissions that are also informative from an implementation perspective, covering different algorithmic designs, modelling assumptions, entropy-coding strategies, and degrees of dataset-specific adaptation.

\subsubsection{G1-V22}

Before compression, the system applies preprocessing transformations to reduce the statistical complexity of the input. These include general-purpose transformations such as BWT, MTF, RLE, delta encoding \cite{suel2018delta}, weighted frequency coding, and x86 BCJ filtering \cite{xzutils_bcj}, as well as image- and audio-oriented transformations such as YCoCg-R color decorrelation, row prediction, stereo mid/side conversion, and wavelet lifting. The applied transform chain is stored in the compressed file header so that decompression can reverse the operations in the correct order. The system also uses a dynamic alphabet, storing only the byte values actually present in the transformed stream, which reduces coding overhead.

The compressor’s machine-learning methodology consists of training two XGBoost-based \cite{chen2016xgboost} selectors to automate pipeline decisions. The first classifier predicts the best transformation chain for a data block, while the second predicts the most suitable compression model after transformation. Both use a 38-dimensional statistical feature vector including entropy measures, run statistics, byte distribution properties, histogram bins, conditional entropies, and LZ77-based features. Training labels are obtained empirically by compressing corpus chunks with all candidate transforms and models, using the smallest output size as the target. These selectors support four heuristic modes—Fast, Normal-Fast, Normal-Accurate, and Accurate—which trade speed for compression quality by either trusting the predictions directly or empirically testing top-ranked combinations on samples before compressing the full file.

The program was implemented in Rust and the source code is freely available at \url{github.com/andreribeiro87/TAI-P1}.

\subsubsection{G2-V3}

The methodology of this compressor separates compression and decompression into four reversible stages: file-type detection, type-specific preprocessing, optional sequence transformation, and adaptive entropy coding with range coding. The detector analyzes at most the first 64 kB of the input and classifies the file into one of eight categories: protein, text, structured text, \textit{random}, float32, generic binary, interleaved, and ELF binary, using features such as zero-order Shannon entropy, alphabet size, printable ASCII ratio, even/odd byte imbalance, float32 exponent-byte patterns, and ELF magic-number detection. Unlike general-purpose compressors such as gzip, bzip, bzip2, xz, or zstd, which normally apply a fixed compression strategy, G2-V3 uses this detection stage to select a specialized reversible pipeline for each type of data. The compressed file header stores the magic value, detected class, original size, and BWT primary index, allowing decompression to reconstruct the exact processing path used during compression. 

Before entropy coding, G2-V3 applies preprocessing transformations designed to expose structure in specific data classes. ELF binaries are processed with an x86 E8/E9 filter, similar in motivation to executable filters used by xz, but integrated here into a type-detection-driven pipeline. Interleaved 16-bit data is split into even and odd byte channels, float32 arrays are transformed using byte-plane splitting, and protein sequences are remapped from their ASCII amino-acid alphabet into a compact 21-symbol range. For selected text-like and ELF data, G2-V3 applies BWT followed by MTF. This stage resembles bzip, which used BWT+MTF with arithmetic coding, more than bzip2, which uses Huffman coding rather than arithmetic coding. However, G2-V3 applies this pipeline selectively through file-type detection.

The final stage combines adaptive statistical models with a range coder. The range coder is the implementation of Daisuke Okanohara. For most file classes, G2-V3 uses an order-1 adaptive model over byte symbols plus EOF, updating frequency and cumulative-frequency tables during both compression and decompression so that the decoder reproduces the same probability estimates without side information. For protein sequences, it uses a PPM-style order-2 model with fallback to order-1 and order-0 contexts through escape symbols, taking advantage of the compact remapped alphabet and local dependencies between amino acids. \textit{Random} files are detected as incompressible and stored directly, avoiding wasted modelling effort, while interleaved files are encoded as independent even and odd streams. 

The program was implemented in C++ and the source code is freely available at \url{https://github.com/diogux/TAI-Zipper}.

\subsubsection{G5-V25}

The methodology of G5-V25 is based on an adaptive arithmetic range coder combined with modular probability models. The system separates entropy coding from modelling, allowing different models to be selected through a common interface. The compressed format stores a magic value, block size, and per-block raw/compressed sizes, supporting independent block compression and parallel execution. G5-V25 belongs to the final context-mixing family; nearby variants such as G5-V23 and G5-V24 use very similar principles and obtain broadly comparable results, mainly differing in block size and parallelisation choices. 

Its main strategy is bitwise context mixing, inspired by PAQ-style compressors. Each byte is encoded with several component models predicting each next bit. These predictions are converted to log-odds, combined with adaptive weights, transformed back into probabilities, and passed to the range coder. The weights are updated by gradient descent after each bit, allowing the compressor to adapt to the most useful contexts. Compared with earlier order-0, order-1, and PPM versions, G5-V25 uses independent discretised context models, an init-only frequency prior, and a word-matching component for text. 

A key feature of the G5-V22 to G5-V25 family is automatic context discovery using Mutual Information. Before encoding, the compressor identifies byte-distance offsets that are most predictive and stores them in the header so the decoder can reproduce the same model. This differs from standard compressors, whose contexts or dictionaries are mostly fixed. G5-V25 therefore emphasizes compression ratio over speed: it is closer to PAQ-like context mixers than to gzip, zstd, or lzma, but is slower because it performs bitwise coding and evaluates multiple models for every bit. 

The program was implemented in C and the source code is freely available at \url{https://github.com/sebastiaoteixeira/G5-V25}.

\subsubsection{G7-V8}

G7-V8 follows a speed-oriented methodology, deliberately moving away from the statistical pipeline used in other versions from the same group. Instead of applying BWT, MTF, adaptive context modelling, and range coding, it uses a simpler LZ77-style dictionary matcher based on a small hash table and a byte-aligned token stream. Its objective is therefore not to maximize compression ratio, but to achieve high throughput with low implementation complexity.

During compression, G7-V8 scans the input in a single pass and searches for repeated byte sequences through hash-based matching. When a match is found, it emits a compact token encoding the match distance and length; otherwise, it outputs literal bytes. This follows the same broad dictionary-compression principle as LZ77-derived compressors such as LZ4 \cite{collet_lz4}. However, G7-V8 additionally performs an explicit initial entropy check to detect near-incompressible input and store such data raw, avoiding unnecessary matching work and limiting expansion on random-like files.

The program was implemented in C++ and the source code is freely available at \url{https://github.com/roldao04/TAI_Proj1}.

\subsubsection{G7-V10}

The methodology of G7-V10 extends the project’s statistical compression line, which is based on the combination of BWT, MTF, adaptive modelling, and range coding. It builds on earlier versions that progressed from order-0 arithmetic coding to order-1 range coding, BWT preprocessing, optimized parallel BWT, and the balanced G7-V5 pipeline, which includes optional zero-run encoding, PPM-style escape handling, and adaptive order-1 modelling.

G7-V10 keeps this statistical foundation but introduces a more flexible block-level selection strategy. Instead of applying a single fixed preprocessing and modelling pipeline to the entire file, each block is evaluated under several candidate configurations, including raw coding, BWT-based coding, LZP-prefiltered coding, x86-filtered coding, and combined LZP+BWT variants. The compressor can also choose between order-1 and order-2 statistical models depending on the block characteristics.

The selected configuration for each block is stored as metadata through transform flags and model indicators, allowing the decompressor to reproduce the exact same processing path. Methodologically, G7-V10 therefore differs from fixed-pipeline compressors such as bzip2 or xz, although it uses familiar transform and entropy-coding components, it applies them adaptively at block level rather than enforcing one uniform compression strategy for the whole input.

Beyond this block-level adaptation, G7-V10 also includes a detector for algorithmically generated data. Before statistical compression, it estimates the Shannon entropy of the input, so if the data appears near-random the compressor tests the hypothesis that it was produced by the \textit{glibc} \texttt{rand()} pseudo-random generator and it brute-forces candidate seeds from 0 to 65535, this while verifying a full-file match. When a seed is found, the entire file is stored as a 14-byte descriptor (model identifier, seed, and original length), from which the decompressor regenerates the data deterministically. This illustrates the distinction between Shannon entropy and Kolmogorov complexity, a file may be statistically incompressible yet have a very short generating program. On the benchmark corpus, File~D is reduced to 14 bytes, while general-purpose compressors store it essentially unchanged.

The program was implemented in C++ and the source code is freely available at \url{https://github.com/roldao04/TAI_Proj1}.

\subsubsection{G9-V1}

The proposed compressor uses a PAQ-style bitwise predictive architecture in which encoder and decoder update the same probability model synchronously from identical deterministic initialization, following the context-mixing principles used in PAQ \cite{mahoney2005adaptive}. Its core units are sparse Gated Linear Network neurons \cite{veness2021gated}, where hashed byte-level $n$-gram contexts of orders 1, 2, 3, 4, 6, 8, and 16, together with partial-byte state, select one context-specific weight per prediction. This preserves GLN adaptivity while reducing inference and learning to constant-time table lookups. Compared with conventional PAQ counters or adaptive maps \cite{mahoney2005adaptive}, the novelty is replacing static context statistics with online-trained, localized GLN weights under symmetric encoder--decoder constraints.

The pipeline combines heterogeneous predictors through cascaded logistic mixers. The ten context models are all GLN neurons: seven capture byte histories at the orders above (orders 3 and 6 only in the Strong profile), while three specialized models handle partial-byte structure, normalized words, and x86 branch-address patterns. Their logits are aggregated by a history mixer, combined with an LZ77 matcher through a structure mixer, and refined by a master mixer, linking neural context prediction with dictionary-style redundancy modelling inspired by Lempel--Ziv compression \cite{ziv1977universal}. After each bit, the global error is propagated backward so each component is reinforced or suppressed according to its contribution. This forms a compact hybrid between neural context mixing and classical dictionary compression.

Determinism and efficiency are ensured through fixed-point arithmetic, a static sigmoid lookup table, compile-time profile specialization, SIMD-accelerated mixer dot products, and optional 16-bit weight quantization. The final probability is corrected with Secondary Symbol Estimation and an Adaptive Probability Map, while preprocessing applies BCJ transformation to x86 data (detected at the beginning of the file using ELF/MZ magic) and bypasses high-entropy 64 kB chunks. Relative to LZ77-family tools \cite{ziv1977universal}, PPM, and PAQ \cite{mahoney2005adaptive}, the main novelty is a deterministic online neural probability engine optimized for bitwise arithmetic coding, motivated by the broader view of compression as probabilistic sequence prediction \cite{solomonoff1964formal,kolmogorov1968three}.

The program was implemented in Rust and the source code is freely available at \url{https://github.com/Mycsina/compaq/tree/bcce088e7cc0b1f966c9d2bf43b1a30dbefeebe4}.

\subsection{External Dataset}
\label{subsec:external_dataset}

We included an external evaluation to test robustness beyond the controlled challenge setting and to assess whether the observed performance trends were specific to the official benchmark or remained valid under broader, more realistic data conditions. For this purpose, in addition to the official benchmark files A--P, we evaluated selected submitted compressors and reference compressors on four large heterogeneous files, labeled Q--T. These files were not part of the challenge leaderboard and were used only for the external generalization analysis.

The external dataset was designed to cover distinct real-world compression regimes. File Q is the human T2T-CHM13v2.0 genome in FASTA format, representing highly repetitive and biologically structured genomic sequence data \cite{Q_T2T_CHM13v2_2022,Q_T2T_CHM13v2_NCBI}. File R is the DBLP bibliographic dump in XML format, containing structured text with repeated tags, author names, titles, venues, and metadata, and therefore representing a favourable case for compressors that exploit textual and markup redundancy \cite{R_DBLP_XML_Dump}. File S is the Caltech-256 image dataset packaged as a TAR archive of JPEG images; because JPEG files are already lossy-compressed, this dataset is expected to provide limited additional compressibility and serves as a difficult case for general-purpose compressors \cite{S_Caltech256_2007}. File T is the OpenStreetMap Portugal extract in PBF format, a compact binary geospatial representation containing nodes, ways, relations, tags, and coordinates; because the PBF format is already space-efficient, this file is expected to be nearly incompressible \cite{T_OSM_Portugal_Geofabrik,T_OSM_PBF_Format}.

For each external source, only the first 250,000,000 symbols were extracted and used in the evaluation. This fixed-size sampling procedure ensured comparability across files while keeping the experiment computationally tractable.

The external evaluation was restricted to the selected submitted compressors and to the reference compressors, with the exception of \texttt{paq8px-1}. This compressor was excluded because its very high computational cost made the full external protocol impractical: according to the same methodology described in the benchmark execution protocol, each compressor had to be run three times for compression and three times for decompression on each of the four 250-million-symbol datasets. Including \texttt{paq8px-1} would therefore have required a disproportionately large amount of execution time relative to the remaining methods. For all included compressors, each file--compressor pair was compressed and decompressed three times, and median compression and decompression times were reported using the same timing convention as in the main benchmark.

\subsection{Computational Environment}

All the benchmark computations were conducted on a laptop running Ubuntu 24.04.2 LTS. The hardware platform was a Dell Latitude 7490 equipped with an Intel Core i7-8650U processor, 16~GB of RAM, and a 512~GB SSD. This environment provided a practical evaluation setup consistent with the objective of assessing compressors under realistic resource constraints rather than specialized high-performance computing conditions.

\section{Results}
\label{sec:results}

\begin{figure}[!h]
\centering
\includegraphics[width=1\textwidth]{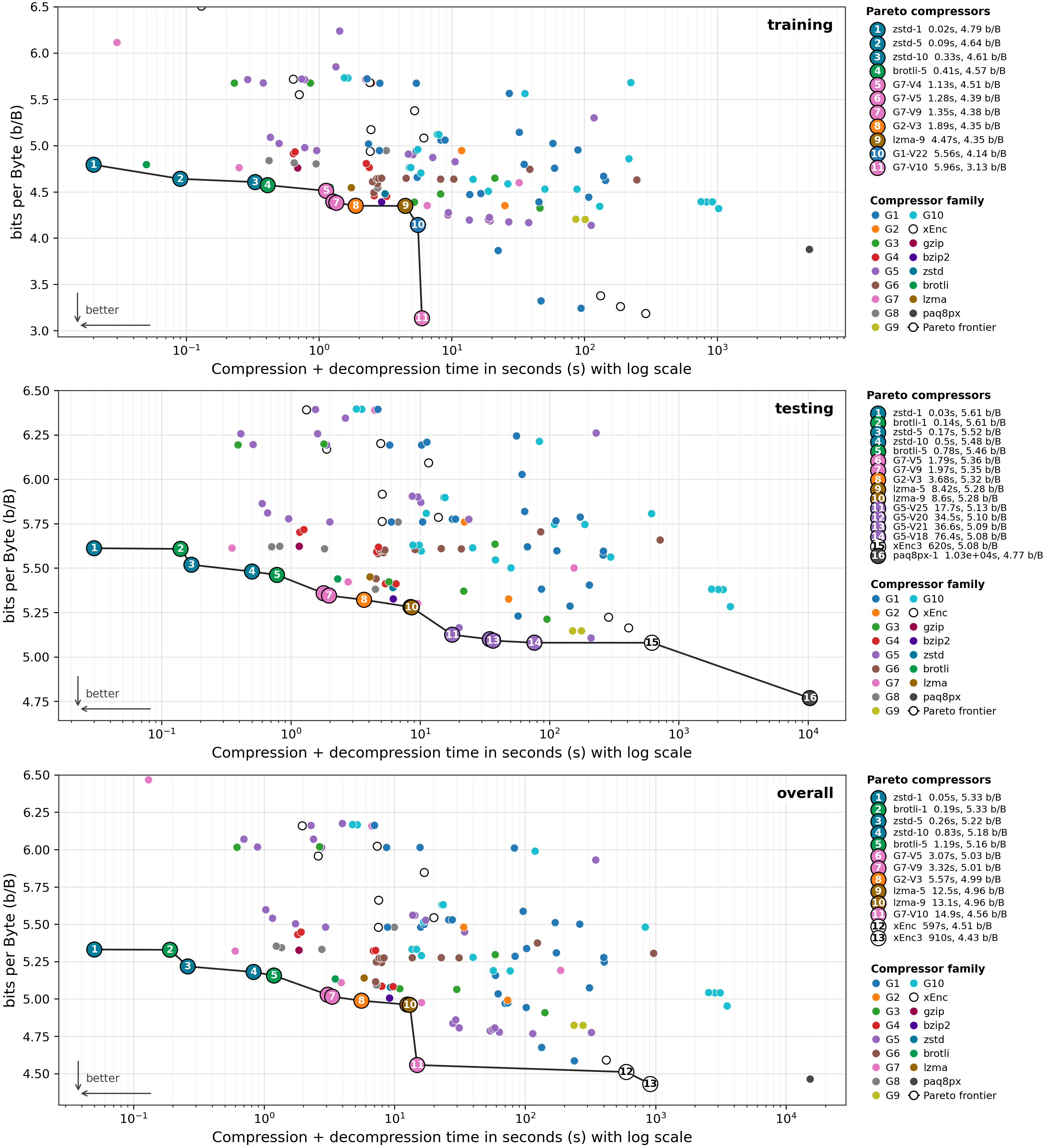}
\caption{Pareto analysis of compressor performance. The three panels show compressor performance for the training datasets, testing datasets, and overall complete dataset collection, from top to bottom. The x-axis shows total compression plus decompression time on a logarithmic scale, and the y-axis shows compressed size in bits per original byte (b/B); lower-left points indicate better performance. Colors denote compressor families. Larger numbered markers and the connecting black line identify Pareto-optimal compressors, for which no other valid compressor is both faster and smaller. The numbered key lists the corresponding compressor names, runtimes, and compressed sizes. For simplicity, the small white circles labeled xEnc also include the ox, ARVIS, fse*, and rc* compressors within the same compressor family; however, only the xEnc versions lie on the Pareto frontier.
}
\label{fig:pareto}
\end{figure}

Figure~\ref{fig:pareto} compares the compressors on the training set, testing set, and complete dataset collection. Each point represents one compressor, with the horizontal axis showing the total execution time, i.e., compression plus decompression time, on a logarithmic scale, and the vertical axis showing the compressed size in bits per original byte. Therefore, compressors closer to the lower-left region provide a better trade-off between speed and compressed size. For readability, the plots focus on the most relevant compression range and omit values above 6.5~b/B, with the complete results available on the challenge website.

Across the three panels, only a small subset of compressors lies on the Pareto frontier, indicating that most methods are dominated by alternatives that are either faster, achieve smaller compressed sizes, or both. The results reveal three main performance profiles. The first corresponds to very fast compressors with comparatively lower compression, including \texttt{zstd-1}, \texttt{brotli-1}, \texttt{zstd-5}, and the challenge submission \texttt{G7-V8}. These methods are attractive when runtime is the main constraint, but they do not target the smallest compressed sizes. The second profile corresponds to balanced compressors that improve compression while maintaining moderate execution times, such as \texttt{zstd-10}, \texttt{brotli-5}, \texttt{G7-V5}, \texttt{G2-V3}, and \texttt{lzma-9}. The third profile corresponds to much slower but highly compressive methods, including \texttt{xEnc3}, \texttt{xEnc}, \texttt{xEnc2}, \texttt{G7-V10}, \texttt{G1-V25}, \texttt{G1-V24}, \texttt{G5-V18}, \texttt{G5-V26}, and the \texttt{G9-V*} variants. These methods substantially reduce the compressed size, but at a much higher computational cost.

A particularly notable result is that \texttt{xEnc3} outperforms \texttt{paq8px-1} in compressed size while requiring substantially less execution time. On the overall dataset, \texttt{xEnc3} produces 21,200,660 compressed bytes, compared with 21,359,591 bytes for \texttt{paq8px-1}, while reducing the total execution time from 15279.74~s to 910.03~s. Other highly compressive submissions, such as \texttt{xEnc}, \texttt{G7-V10}, \texttt{G1-V25}, \texttt{xEnc2}, and \texttt{G1-V24}, do not surpass \texttt{paq8px-1} in compressed size on the overall dataset, but remain close while requiring substantially less time. In particular, \texttt{G7-V10} reaches 21,811,426 compressed bytes in only 14.88~s, illustrating a markedly different speed--compression trade-off.

\texttt{G7-V10} is another notable case. This compressor appears as a strong Pareto point because it is highly optimized for one specific file, namely file~\texttt{D}. For this file, the generation process was effectively identified by the authors: the data were produced by a linear congruential generator (LCG), and the corresponding parameters could be exploited to encode the file extremely compactly. However, this behaviour is dataset-specific rather than generally robust. In particular, the same strategy does not transfer when the generation seed changes, as observed for file~\texttt{L}. Therefore, although \texttt{G7-V10} is Pareto-optimal in the plots and ranks highly in terms of compressed size, its performance should be interpreted as a specialized reconstruction of a known generative mechanism where the broadly applicable compression performance is more modest.

Most of the compressors submitted as group variants, such as \texttt{G1-V*}, \texttt{G3-V*}, \texttt{G5-V*}, \texttt{G7-V*}, \texttt{G9-V*}, and \texttt{G10-V*}, as well as the \texttt{xEnc} family, rely on arithmetic or range-coding mechanisms. These approaches are expected to be slower than compressors based on faster entropy coders, such as finite-state entropy (FSE) \cite{collet2013fse,duda2013ans}, but they can provide higher compression ratios because they support more adaptive and fine-grained probabilistic modelling. This is important beyond operational compression. Such compressors can also be used for direct compression-based analysis of the files, for example to estimate complexity profiles or to compare structural regularities across datasets. Therefore, developing these compressors also contributes to improving compression-based analysis tools, even when their runtime is not competitive with the fastest general-purpose methods.

The Weissman score provides a complementary view by combining compression effectiveness and runtime, using \texttt{gzip} as the reference compressor. Table~\ref{tab:weissman_top20} reports the top-20 compressors according to this score on the overall dataset. The ranking strongly favours compressors that are very fast while still achieving reasonable compression. Consequently, \texttt{zstd-1} obtains the highest score, followed by fast methods such as \texttt{G7-V8}, \texttt{brotli-1}, and \texttt{zstd-5}. Several submitted compressors, including \texttt{G7-V7}, \texttt{G3-V5}, and multiple \texttt{G5-V*} variants, also appear among the best Weissman-score results. This indicates that some challenge submissions occupy an effective speed--compression compromise, even when they are not the best compressors in absolute compressed size.

\begin{table}[!h]
\centering
\caption{Top-20 compressors by Weissman score on the overall dataset, using \texttt{gzip} as reference. Compression ratio is reported as original size divided by compressed size. Times are reported as compression time, decompression time, and total time.}
\label{tab:weissman_top20}
{\small
\setlength{\tabcolsep}{3.5pt}
\renewcommand{\arraystretch}{0.92}
\begin{tabular}{r l r r r r r r}
\hline
Rank & Compressor & Weissman score & Comp. ratio & b/B & $t_{\mathrm{comp}}$ (s) & $t_{\mathrm{decomp}}$ (s) & $t_{\mathrm{total}}$ (s) \\
\hline
1  & \texttt{zstd-1}   & 21.446 & 1.501 & 5.331 & 0.05 & 0.00 & 0.05 \\
2  & \texttt{G7-V8}    & 7.057  & 1.237 & 6.467 & 0.08 & 0.05 & 0.13 \\
3  & \texttt{brotli-1} & 6.017  & 1.501 & 5.329 & 0.08 & 0.11 & 0.19 \\
4  & \texttt{zstd-5}   & 4.626  & 1.533 & 5.217 & 0.26 & 0.00 & 0.26 \\
5  & \texttt{fse2}     & 2.462  & 1.213 & 6.595 & 0.26 & 0.15 & 0.41 \\
6  & \texttt{ARVIS}    & 2.395  & 1.106 & 7.230 & 0.28 & 0.10 & 0.38 \\
7  & \texttt{G7-V7}    & 2.231  & 1.504 & 5.321 & 0.39 & 0.21 & 0.60 \\
8  & \texttt{G3-V5}    & 1.922  & 1.330 & 6.015 & 0.26 & 0.36 & 0.62 \\
9  & \texttt{zstd-10}  & 1.782  & 1.544 & 5.180 & 0.82 & 0.01 & 0.83 \\
10 & \texttt{G5-V10}   & 1.732  & 1.318 & 6.070 & 0.21 & 0.49 & 0.70 \\
11 & \texttt{G5-V11}   & 1.456  & 1.330 & 6.017 & 0.41 & 0.48 & 0.89 \\
12 & \texttt{G5-V12}   & 1.408  & 1.429 & 5.597 & 0.51 & 0.52 & 1.03 \\
13 & \texttt{brotli-5} & 1.380  & 1.552 & 5.156 & 1.09 & 0.10 & 1.19 \\
14 & \texttt{G5-V13}   & 1.307  & 1.444 & 5.540 & 0.54 & 0.62 & 1.16 \\
15 & \texttt{G8-V4}    & 1.292  & 1.494 & 5.353 & 0.74 & 0.50 & 1.24 \\
16 & \texttt{G8-V3}    & 1.216  & 1.497 & 5.343 & 0.69 & 0.67 & 1.36 \\
17 & \texttt{G5-V14}   & 1.006  & 1.454 & 5.503 & 0.79 & 0.95 & 1.74 \\
18 & \texttt{gzip}     & 1.000  & 1.502 & 5.326 & 1.66 & 0.19 & 1.85 \\
19 & \texttt{G4-V8}    & 0.994  & 1.473 & 5.431 & 0.79 & 1.02 & 1.81 \\
20 & \texttt{G4-V4}    & 0.956  & 1.469 & 5.447 & 0.92 & 1.00 & 1.92 \\
\hline
\end{tabular}
}
\end{table}

Table~\ref{tab:bpb_top20} ranks the compressors only by compressed size, expressed in bits per original byte. This ranking highlights a different set of methods. The best results are achieved by \texttt{xEnc3}, \texttt{paq8px-1}, \texttt{xEnc}, \texttt{G7-V10}, \texttt{G1-V25}, \texttt{xEnc2}, and \texttt{G1-V24}. These compressors are clearly oriented toward maximizing compression ratio, equivalently minimizing compressed size, rather than minimizing runtime. This ranking is particularly relevant for medium- to long-term storage, where compression and decompression may be performed infrequently and the reduction in storage cost can justify substantially longer execution times.

\begin{table}[!h]
\centering
\caption{Top-20 compressors by compressed size on the overall dataset. Lower b/B indicates stronger compression. Compression ratio is reported as original size divided by compressed size. Times are reported as compression time, decompression time, and total time.}
\label{tab:bpb_top20}
{\small
\setlength{\tabcolsep}{3.5pt}
\renewcommand{\arraystretch}{0.92}
\begin{tabular}{r l r r r r r r}
\hline
Rank & Compressor & Compressed bytes & Comp. ratio & b/B & $t_{\mathrm{comp}}$ (s) & $t_{\mathrm{decomp}}$ (s) & $t_{\mathrm{total}}$ (s) \\
\hline
1  & \texttt{xEnc3}    & 21,200,660 & 1.806 & 4.430 & 452.76  & 457.27  & 910.03 \\
2  & \texttt{paq8px-1} & 21,359,591 & 1.793 & 4.463 & 7636.61 & 7643.13 & 15279.74 \\
3  & \texttt{xEnc}     & 21,586,841 & 1.774 & 4.510 & 294.95  & 301.62  & 596.57 \\
4  & \texttt{G7-V10}   & 21,811,426 & 1.755 & 4.557 & 12.37   & 2.51    & 14.88 \\
5  & \texttt{G1-V25}   & 21,942,639 & 1.745 & 4.585 & 161.89  & 76.53   & 238.42 \\
6  & \texttt{xEnc2}    & 21,967,574 & 1.743 & 4.590 & 207.39  & 211.62  & 419.01 \\
7  & \texttt{G1-V24}   & 22,373,823 & 1.711 & 4.675 & 87.93   & 45.90   & 133.83 \\
8  & \texttt{G5-V18}   & 22,815,305 & 1.678 & 4.767 & 57.08   & 57.38   & 114.46 \\
9  & \texttt{G5-V26}   & 22,850,170 & 1.676 & 4.774 & 293.12  & 29.03   & 322.15 \\
10 & \texttt{G5-V21}   & 22,863,782 & 1.675 & 4.777 & 32.04   & 31.47   & 63.51 \\
11 & \texttt{G5-V20}   & 22,905,321 & 1.672 & 4.786 & 27.14   & 26.81   & 53.95 \\
12 & \texttt{G5-V19}   & 22,957,666 & 1.668 & 4.797 & 28.51   & 27.87   & 56.38 \\
13 & \texttt{G5-V22}   & 22,998,905 & 1.665 & 4.805 & 29.30   & 29.29   & 58.59 \\
14 & \texttt{G5-V25}   & 23,001,857 & 1.665 & 4.806 & 15.63   & 15.61   & 31.24 \\
15 & \texttt{G9-V4}    & 23,081,709 & 1.659 & 4.823 & 136.97  & 139.97  & 276.94 \\
16 & \texttt{G9-V5}    & 23,081,709 & 1.659 & 4.823 & 137.28  & 140.52  & 277.80 \\
17 & \texttt{G9-V6}    & 23,081,709 & 1.659 & 4.823 & 137.56  & 140.29  & 277.85 \\
18 & \texttt{G9-V1}    & 23,084,858 & 1.659 & 4.823 & 117.78  & 118.79  & 236.57 \\
19 & \texttt{G9-V2}    & 23,084,858 & 1.659 & 4.823 & 117.83  & 118.99  & 236.82 \\
20 & \texttt{G9-V3}    & 23,084,858 & 1.659 & 4.823 & 118.28  & 118.81  & 237.09 \\
\hline
\end{tabular}
}
\end{table}

Taken together, the Pareto frontiers, Weissman-score ranking, and b/B ranking show that there is no single universally best compressor. Fast compressors such as \texttt{zstd-1}, \texttt{brotli-1}, \texttt{zstd-5}, and \texttt{G7-V8} are preferable when execution time is the dominant constraint. Balanced methods such as \texttt{zstd-10}, \texttt{brotli-5}, \texttt{G7-V5}, \texttt{G2-V3}, and \texttt{lzma-9} provide stronger compression at moderate runtime cost. In contrast, compressors such as \texttt{xEnc3}, \texttt{xEnc}, \texttt{xEnc2}, \texttt{G7-V10}, \texttt{G1-V25}, \texttt{G1-V24}, \texttt{G5-V18}, \texttt{G5-V26}, \texttt{G5-V21}, and the \texttt{G9-V*} variants are preferable when the main objective is to minimize compressed size, even at substantial computational cost. The choice of compressor therefore depends on the intended use case: operational compression favours speed-balanced methods, medium- to long-term storage favours high compression ratios, and compression-based data analysis or complexity estimation may benefit from slower, more adaptive models with stronger compression capability.

Overall, the training, testing, and complete-data panels show similar qualitative trends: fast compressors dominate the low-runtime region, while more complex compressors are required to reach the lowest compressed sizes. Nevertheless, the exact Pareto-optimal set changes between the training and testing panels, showing that compressor ranking is sensitive to the dataset subset. This is especially important for deployment: a compressor that is optimal on the training set may rely on properties that do not generalize to the testing set. The overall panel therefore provides the most balanced view, highlighting the compressors that offer the best global trade-offs between runtime and compression effectiveness across all files.


\subsection{Binary Similarity Analysis}
\label{subsec:ncd_results}

\begin{figure}[!h]
\centering
\includegraphics[width=1\textwidth]{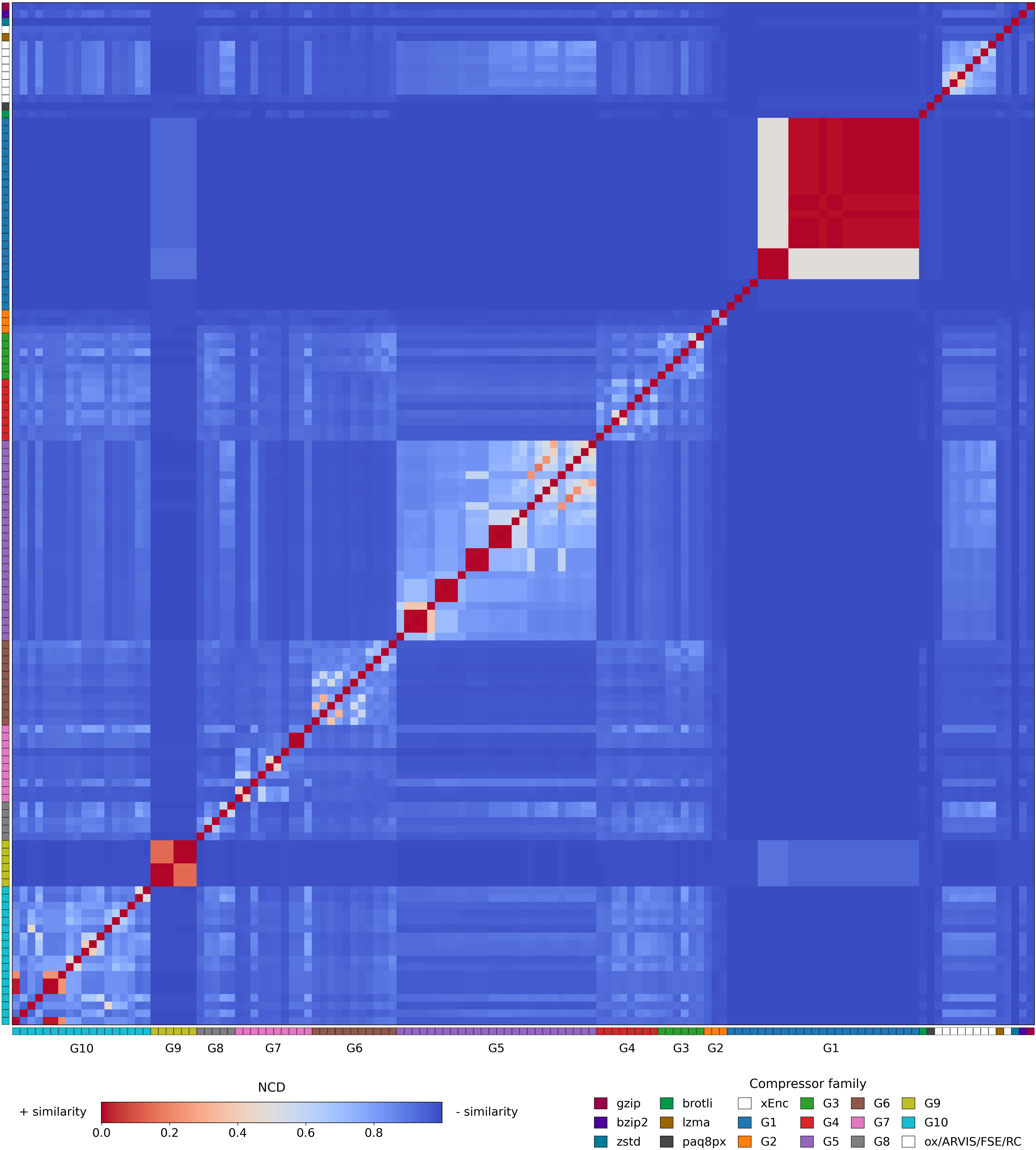}
\caption{Normalized Compression Distance (NCD) heatmap computed between compressor binaries using \texttt{zstd -15}. Each matrix entry represents the pairwise NCD between two compressor executables. Lower values indicate greater binary-level similarity, whereas values close to one indicate limited shared compressible structure. Colored annotations identify compressor families or submission groups. Within each group, compressor versions \texttt{V*} are ordered in ascending order from top to bottom and in descending order from left to right.}
\label{fig:ncd}
\end{figure}

Figure~\ref{fig:ncd} shows the NCD heatmap obtained from the evaluated compressor binaries. Several clear low-distance blocks are visible, indicating strong binary-level similarity among subsets of submissions. These blocks occur mainly within successive versions from the same group, which is consistent with iterative development where most of the implementation is preserved and only modelling parameters, context definitions, coding details, or auxiliary routines are modified.

The strongest similarity patterns were observed among near-identical or closely related version families. For example, the early \texttt{G1-V*} submissions form a compact block with very low NCD values, indicating substantial binary continuity across versions. Similar behaviour is visible for the \texttt{G7-V2}/\texttt{G7-V3} pair, for subsets of the \texttt{G9-V*} variants, and for several \texttt{G10-V*} variants. These results suggest that many submitted compressors evolved through incremental refinements rather than complete rewrites.

A broader visible pattern is also observed between the \texttt{G1} and \texttt{G9} compressor families. Although these compressors correspond to different algorithmic designs, they show a degree of binary-level relatedness that is not observed as strongly with most other submitted tools. This pattern is likely correlated with the implementation language: both \texttt{G1} and \texttt{G9} were implemented in Rust, whereas the remaining submitted compressors and reference tools were implemented in C or C++. Therefore, part of the observed NCD similarity may reflect shared language-level, compiler, runtime, standard-library, or build-system structure rather than direct algorithmic or source code similarity. This is an important caveat when interpreting NCD over executable binaries, because the distance captures all compressible regularities in the binary, not only those associated with the compression algorithm itself.

In contrast, standard reference compressors and independent implementations generally show NCD values close to one when compared with most student submissions. This indicates limited shared binary-level structure and supports the interpretation that these tools are implementation-wise distinct from the submitted compressors. Intermediate NCD values appear between some compressors from different groups or distant versions, suggesting partial reuse of components, similar implementation templates, common build artifacts, convergent implementation structure, or common effects introduced by the programming language and compilation toolchain.

Overall, the NCD analysis complements the compression-performance results by showing that the submitted compressors were diverse not only in compression ratio and runtime, but also in binary-level relatedness. The heatmap also helps distinguish genuine algorithmic diversity from families of closely related variants. At the same time, the visible association between the Rust-based \texttt{G1} and \texttt{G9} families shows that binary-level NCD should be interpreted carefully: low-distance patterns can arise from common implementation languages or toolchains as well as from shared compressor design. This distinction is important when interpreting leaderboard rankings based on many successive submissions from the same development line.

\subsection{External Analysis on Selected Compressors}
\label{subsec:selected}

Table~\ref{tab:external_results} reports the performance of the reference compressors and the selected submitted compressors on the external datasets Q--T. All evaluated compressors were verified to be lossless on these larger external files, each containing 250,000,000 symbols, confirming that the implementations were not only valid on the official benchmark but also robust when applied to substantially larger heterogeneous datasets.

\begin{table*}[!ht]
\centering
\caption{External dataset results for the selected compressors and reference compressors. Each external file contains the first 250,000,000 symbols extracted from the corresponding original source. For each dataset, the table reports bits per original byte (b/B), median compression time, and median decompression time. Times are reported in seconds. Lower b/B values indicate stronger compression. Bold values indicate the minimum value in each column.}
\label{tab:external_results}
{\scriptsize
\setlength{\tabcolsep}{3.2pt}
\renewcommand{\arraystretch}{1.08}
\resizebox{\textwidth}{!}{%
\begin{tabular}{l rrr rrr rrr rrr rrr}
\toprule
\multirow{2}{*}{Compressor}
& \multicolumn{3}{c}{Q}
& \multicolumn{3}{c}{R}
& \multicolumn{3}{c}{S}
& \multicolumn{3}{c}{T}
& \multicolumn{3}{c}{Overall} \\
\cmidrule(lr){2-4}
\cmidrule(lr){5-7}
\cmidrule(lr){8-10}
\cmidrule(lr){11-13}
\cmidrule(lr){14-16}
& \multicolumn{1}{r}{b/B} & \multicolumn{1}{r}{$t_c$} & \multicolumn{1}{r}{$t_d$}
& \multicolumn{1}{r}{b/B} & \multicolumn{1}{r}{$t_c$} & \multicolumn{1}{r}{$t_d$}
& \multicolumn{1}{r}{b/B} & \multicolumn{1}{r}{$t_c$} & \multicolumn{1}{r}{$t_d$}
& \multicolumn{1}{r}{b/B} & \multicolumn{1}{r}{$t_c$} & \multicolumn{1}{r}{$t_d$}
& \multicolumn{1}{r}{b/B} & \multicolumn{1}{r}{$t_c$} & \multicolumn{1}{r}{$t_d$} \\
\midrule
\texttt{gzip}
& 2.404 & 41.0 & 1.4
& 1.633 & 5.2 & 1.4
& 7.635 & 7.8 & 1.6
& 7.988 & 7.0 & 1.6
& 4.915 & 61.0 & 6.0 \\

\texttt{bzip2}
& 2.111 & 39.0 & 13.1
& 0.998 & 56.3 & 8.6
& 7.650 & 47.3 & 24.6
& 8.021 & 49.6 & 25.9
& 4.695 & 192.2 & 72.2 \\

\texttt{zstd-1}
& 2.499 & \textbf{0.6} & \textbf{0.4}
& 1.675 & \textbf{0.3} & \textbf{0.2}
& 7.640 & \textbf{0.3} & \textbf{0.1}
& 8.000 & \textbf{0.2} & \textbf{0.1}
& 4.954 & \textbf{1.4} & \textbf{0.9} \\

\texttt{zstd-5}
& 2.416 & 1.6 & 0.5
& 1.310 & 1.0 & 0.3
& 7.593 & 0.7 & 0.1
& 8.000 & 0.3 & 0.1
& 4.830 & 3.7 & 1.0 \\

\texttt{zstd-10}
& 2.291 & 5.8 & 0.4
& 1.151 & 3.5 & 0.3
& 7.576 & 2.2 & 0.1
& 8.000 & 1.1 & 0.1
& 4.754 & 12.7 & 1.0 \\

\texttt{zstd-18}
& 1.976 & 137.5 & 0.5
& 1.032 & 48.5 & 0.3
& 7.556 & 33.9 & 0.2
& 8.000 & 35.0 & 0.1
& 4.641 & 254.9 & 1.2 \\

\texttt{ox}
& 3.020 & 2.5 & 6.8
& 5.309 & 3.0 & 8.8
& 7.937 & 2.7 & 13.2
& 8.000 & 2.7 & 13.6
& 6.066 & 11.0 & 42.4 \\

\texttt{lzma-1}
& 2.387 & 16.4 & 3.1
& 1.275 & 8.8 & 1.5
& 7.623 & 45.4 & 6.4
& 8.071 & 51.5 & 6.6
& 4.839 & 122.1 & 17.6 \\

\texttt{lzma-5}
& 1.927 & 245.9 & 2.2
& 1.065 & 67.5 & 1.3
& 7.590 & 107.5 & 6.3
& 8.063 & 112.9 & 6.6
& 4.661 & 533.8 & 16.5 \\

\texttt{lzma-9}
& 1.874 & 384.2 & 3.1
& 0.938 & 114.6 & 1.3
& 7.589 & 139.2 & 6.5
& 8.063 & 147.0 & 6.8
& 4.616 & 785.0 & 17.7 \\

\texttt{ARVIS}
& 3.217 & 2.0 & 2.2
& 5.638 & 2.9 & 2.9
& 8.000 & 1.9 & 0.4
& 8.000 & 1.9 & 0.4
& 6.214 & 8.8 & 6.0 \\

\texttt{fse2}
& 2.918 & 2.0 & 2.1
& 5.499 & 2.9 & 2.8
& 8.003 & 1.7 & 0.2
& 8.003 & 1.7 & 0.2
& 6.106 & 8.3 & 5.3 \\

\texttt{fseo1}
& 2.137 & 2.6 & 2.1
& 3.781 & 4.8 & 3.0
& 7.938 & 36.8 & 1.4
& 8.003 & 35.2 & 0.3
& 5.465 & 79.4 & 6.9 \\

\texttt{rc32}
& 2.874 & 14.7 & 15.7
& 5.269 & 20.9 & 24.4
& 7.890 & 25.7 & 33.0
& 8.001 & 25.6 & 33.5
& 6.008 & 86.9 & 106.7 \\

\texttt{rcox1}
& 2.062 & 13.6 & 16.8
& 3.323 & 17.8 & 23.4
& 7.736 & 25.5 & 36.5
& 7.992 & 25.4 & 37.0
& 5.278 & 82.3 & 113.7 \\

\texttt{rcox1b}
& 2.030 & 14.0 & 17.0
& 2.187 & 16.5 & 22.8
& 7.702 & 71.6 & 103.5
& 7.987 & 73.9 & 108.1
& 4.977 & 175.9 & 251.3 \\

\texttt{rcox2}
& 2.061 & 13.4 & 16.7
& 3.315 & 17.6 & 23.0
& 7.736 & 25.4 & 36.8
& 7.992 & 25.8 & 37.5
& 5.276 & 82.1 & 114.0 \\

\texttt{rcmix}
& 2.092 & 19.6 & 27.3
& 2.444 & 23.4 & 37.7
& 7.664 & 82.6 & 127.7
& \textbf{7.981} & 85.8 & 139.0
& 5.045 & 211.4 & 331.7 \\

\texttt{brotli-1}
& 2.606 & 1.3 & 0.8
& 1.759 & 0.9 & 0.7
& 7.647 & 0.7 & 0.7
& 8.000 & 0.3 & 0.3
& 5.003 & 3.1 & 2.5 \\

\texttt{brotli-5}
& 2.298 & 11.5 & 0.7
& 1.178 & 4.9 & 0.5
& 7.564 & 3.4 & 0.9
& 8.000 & 1.8 & 0.2
& 4.760 & 21.6 & 2.4 \\

\texttt{brotli-9}
& 2.171 & 87.3 & 0.7
& 1.055 & 51.3 & 0.6
& \textbf{7.545} & 130.6 & 0.9
& 8.000 & 70.5 & 0.2
& 4.693 & 339.7 & 2.4 \\

\midrule

\texttt{G1-V22}
& 2.149 & 30.0 & 4.4
& 5.271 & 38.2 & 5.7
& 9.626 & 35.9 & 10.6
& 8.618 & 5.1 & 8.7
& 6.416 & 109.1 & 29.3 \\

\texttt{G2-V3}
& 1.934 & 162.5 & 53.5
& \textbf{0.822} & 146.8 & 44.3
& 7.730 & 10.4 & 21.2
& 8.000 & 0.6 & 0.6
& 4.622 & 320.4 & 119.5 \\

\texttt{G5-V25}
& 1.994 & 50.7 & 52.4
& 1.332 & 53.6 & 55.4
& 7.562 & 128.2 & 126.0
& 8.011 & 134.8 & 132.2
& 4.725 & 367.3 & 365.9 \\

\texttt{G7-V8}
& 5.214 & 1.0 & 1.0
& 2.449 & 0.9 & 0.8
& 7.678 & 0.8 & 0.6
& 8.000 & 0.5 & 0.5
& 5.835 & 3.3 & 2.8 \\

\texttt{G7-V10}
& 1.937 & 36.9 & 4.2
& 1.015 & 33.3 & 3.3
& 7.937 & 3.1 & 2.5
& 8.000 & 0.7 & 0.5
& 4.722 & 73.9 & 10.5 \\

\texttt{G9-V1}
& \textbf{1.847} & 898.3 & 818.7
& 0.839 & 854.8 & 796.1
& 7.574 & 569.6 & 588.2
& 8.000 & 0.8 & 0.7
& \textbf{4.565} & 2323.5 & 2203.7 \\

\bottomrule
\end{tabular}%
}
}
\end{table*}

The results show that several selected compressors are competitive with, and in specific cases outperform, widely used state-of-the-art general-purpose compressors. The most notable case is \texttt{G9-V1}, which obtains the best compression on file Q, reaching 1.847~b/B,  improving on \texttt{lzma-9} at 1.874~b/B. It also achieves the best overall compression across the four external datasets, with 4.565~b/B, outperforming the strongest compressors in the benchmark, including \texttt{lzma-9}, \texttt{zstd-18}, and \texttt{brotli-9}. This result indicates that the PAQ-style neural/context-mixing strategy of \texttt{G9-V1} generalizes well to external data, although at a high computational cost.

The compressor \texttt{G2-V3} is particularly effective on file R, where it obtains the best result in the table, with 0.822~b/B. This is substantially better than strong established methods such as \texttt{lzma-9}, \texttt{bzip2}, \texttt{zstd-18}, and \texttt{brotli-9}, suggesting that its type-aware preprocessing and adaptive modelling are especially well suited to highly structured textual or markup-like data. In the overall external evaluation, \texttt{G2-V3} also achieves the third-best compressed size, with 4.622~b/B, slightly better than \texttt{zstd-18} and \texttt{brotli-9}. Although it is slower than these faster reference compressors, it requires less total execution time than \texttt{lzma-9} (less than half), providing a favourable trade-off between compression effectiveness and computational cost.

The compressor \texttt{G7-V10} also shows competitive behaviour. While it is not the best compressor in absolute compressed size, it obtains 4.722~b/B overall, outperforming several widely used references such as \texttt{gzip}, \texttt{bzip2}, \texttt{zstd-1}, \texttt{zstd-5}, \texttt{zstd-10}, \texttt{lzma-1}, \texttt{brotli-1}, and \texttt{brotli-5}. Its performance on file R is also strong, with 1.015~b/B, improving over \texttt{zstd-18} and \texttt{brotli-9}. This confirms that adaptive block-level selection of transforms and models can provide useful generalization beyond the original challenge files.

The results for \texttt{G5-V25} show a different profile. It is not the best method overall, but it remains competitive on the less compressible image-archive dataset S, where it obtains 7.562~b/B, close to the best reference results. This behaviour is consistent with a context-mixing compressor that can still extract limited residual redundancy from data that are already partially compressed. In contrast, \texttt{G7-V8} is clearly speed-oriented: it gives weaker compression ratios, but its compression and decompression times are among the fastest submitted methods, illustrating a distinct design point in the speed--compression trade-off.

Dataset T appears to be nearly \textit{incompressible} for almost all compressors, with most values close to 8~b/B. This is consistent with the nature of compact binary formats and indicates that, on this file, the main advantage comes not from modelling power but from avoiding unnecessary expansion and computation. In this case, several compressors, including \texttt{G2-V3}, \texttt{G7-V8}, \texttt{G7-V10}, and \texttt{G9-V1}, remain close to the incompressibility limit.

Overall, the external evaluation confirms that the submitted benchmark compressors are not merely tuned to the official challenge data. Several of the (submitted) selected methods outperform widely used state-of-the-art general-purpose compressors in specific regimes: \texttt{G9-V1} on genomic-like data and overall compressed size, \texttt{G2-V3} on structured text, and \texttt{G7-V10} as a competitive adaptive transform-based compressor. These results highlight the scientific value of the benchmark: it provides a controlled but diverse framework for testing new compression ideas under practical constraints, while the external datasets further assess generalization.

The importance of this benchmark goes beyond the numerical rankings. By making the benchmark resources, submitted compressors, and source code available, the challenge provides the community with reusable implementations, new modelling strategies, and experimental baselines that can support future work in lossless compression. The developed methods may help improve compression efficiency, reduce storage and transmission costs, and support more sustainable data-intensive computing. In this sense, the benchmark contributes not only to algorithmic information theory and data compression research, but also to broader sustainability goals, including more efficient digital infrastructure, responsible resource use, and reductions in the energy and environmental costs associated with large-scale data storage and communication.


\section{Conclusion}
\label{sec:conclusion}

This work analyzed the 2026 Algorithmic Information Theory Data Compression Challenge as a practical benchmark for general-purpose lossless compression under realistic constraints. The challenge combined a heterogeneous dataset collection, a hidden testing partition during the competition, decompressor-size and memory limits, and a common evaluation protocol, thereby encouraging compressors that generalize beyond a single data type or public training set.

The results show that no single compressor dominates across all criteria. Fast compressors such as \texttt{zstd-1}, \texttt{brotli-1}, and \texttt{G7-V8} provide the best choices when runtime is the primary constraint, whereas modelling-intensive compressors such as \texttt{xEnc3}, \texttt{xEnc}, \texttt{G1-V25}, \texttt{G1-V24}, \texttt{G5-V18}, and several \texttt{G9-V*} variants achieve smaller compressed sizes at substantially higher computational cost. The Pareto-frontier analysis makes these trade-offs explicit, while the Weissman score provides a complementary view that favours compressors with favourable speed--compression balance.

The analysis also shows that at least one submitted compressor, \texttt{xEnc3}, can outperform the strong reference method \texttt{paq8px-1} in compressed size while requiring considerably less execution time. Other submitted compressors approach the compressed size of \texttt{paq8px-1} while offering substantially lower runtime, highlighting alternative speed--compression trade-offs. At the same time, the comparison between training and testing results highlights the importance of hidden evaluation data, since some high-performing methods may exploit dataset-specific structure rather than broadly generalizable compression principles. The NCD-based binary similarity analysis further complements the performance evaluation by revealing which submissions correspond to closely related implementation families and which appear more independent at the binary level.

Overall, the challenge reinforces the importance of modelling in lossless compression and illustrates the continued relevance of compression as both an engineering problem and an empirical tool for studying structure in data. By making the benchmark, leaderboard, and submitted binaries publicly available, the challenge provides a useful resource for future work on compressor design, benchmarking methodology, and compression-based analysis.

\section*{Data and Code Availability}

The benchmark resources, public leaderboard, and compressor binaries are available from the challenge website at \url{https://aitdcc.github.io/}. The article reports results obtained from the public benchmark outputs and associated evaluation scripts.

\section*{Competing interests}
The authors declare no competing interests.




\bibliographystyle{abbrv}

\bibliography{references.bib}


\end{document}